# Electrical barriers and their elimination by tuning (Zn,Mg)O composition in Cu(In,Ga)S$_2$: Systematic approach to achieve over 14% power conversion efficiency

Mohit Sood*[1], Poorani Gnanasambandan[1,2], Damilola Adeleye[1], Sudhanshu Shukla[1], Noureddine Adjeroud[2], Renaud Leturcq[2], Susanne Siebentritt[1]

(E-mail address: mohit.sood@uni.lu)

[1]*Laboratory for Photovoltaics, Department of Physics and Materials Science, University of Luxembourg, Belvaux, L-4422, Luxembourg*

[2]*Material Research and Technology department, Luxembourg Institute of Science and Technology, Belvaux, L-4422, Luxembourg*

## Abstract

Traditional CdS buffer layer in selenium-free Cu(In,Ga)S$_2$ solar cells leads to reduced open-circuit voltage because of a negative conduction band offset at the Cu(In,Ga)S$_2$/CdS interface. Reducing this loss necessitates the substitution of CdS by an alternative buffer layer. However, the substitute buffer layer may introduce electrical barriers in the device due to unfavorable band alignment at the other interfaces such as between buffer/ZnO i-layer. This study aims to reduce interface recombinations and eliminate electrical barriers in Cu(In,Ga)S$_2$ solar cells using a combination of Zn$_{1-x}$Mg$_x$O and Al-doped Zn$_{1-x}$Mg$_x$O buffer and i-layer combination deposited using atomic layer deposition and magnetron sputtering, respectively. The devices prepared with these layers are characterized by current-voltage and photoluminescence measurements. Numerical simulations are performed to comprehend the influence of electrical barriers on the device characteristics. An optimal composition of $Zn_{1-x}Mg_xO$ x = 0.27 is identified for a suitable conduction band alignment with $Cu(In,Ga)S_2$ with a bandgap of ~1.6 eV, suppressing interface recombination and avoiding barriers. Optimized buffer composition together with a suitable i-layer led to a device with 14 % efficiency and an open-circuit voltage of 943 mV. A comparison of optoelectronic measurements for devices prepared with ZnO and Al:(Zn,Mg)O shows the necessity to replace the ZnO i-layer with Al:(Zn,Mg)O i-layer for a high-efficiency device.

*corresponding author: mohit.sood@uni.lu

# Introduction

Buffer and window layers are essential for p-n junction formation and charge carrier separation in heterojunction chalcopyrite solar cells. Traditionally, cadmium sulfide (CdS) buffer layer in combination with intrinsic zinc oxide (ZnO) and aluminum-doped zinc oxide (Al:ZnO) window layer is used as optimum partners for high-efficiency copper indium gallium sulfoselenide Cu(In,Ga)(S,Se)$_2$ devices.[1-6] One of the reasons for this combination is that these layers have a rather low conduction band offset,[7, 8] and consequently, there are no electrical barriers at the interfaces in between these layers. Cu(In,Ga)S$_2$ (CIGSu), the selenium-free counterpart of CIGSe absorber material, has immense potential to be deployed as a top cell in tandem devices together with a Si or CIGSe bottom cell, owing to its tunable direct bandgap (E$_G$) between 1.54 eV (CuInS$_2$) to 2.53eV (CuGaS$_2$)[9]. However, the Cu-deficient CIGSu devices fabricated with the traditional CdS buffer layer suffer from dominant interface recombination, due to negative conduction band offset or a 'cliff' at the CIGSu/CdS interface.[10] This leads to a difference between the maximum achievable open-circuit voltage given by the quasi-Fermi level splitting (qFLs) and the actual open-circuit voltage in the device (V$_{OC}$).[11] Higher power conversion efficiency (PCE) aspirations require the substitution of the CdS buffer layer with alternative buffer layers that possess a higher conduction band minimum (CBM) energy value than CdS. The zinc oxy-sulfide Zn(O,S), zinc magnesium oxide ((Zn,Mg)O), buffers are particularly interesting, as it is possible to modify the bandgap and the CBM energy of the films by tuning the O/S ratio and the Mg content in the films, respectively.[12-14] As a matter of fact, the record PCE of 15.5% has been achieved by replacing the CdS/i-ZnO buffer i-layer stack with a bilayer (Zn,Mg)O buffer stack.[15] Here, we present qFLs and V$_{OC}$ measurements to better understand the losses, together with a model of how replacing these layers improves the device performance.

We recently demonstrated a CIGSu device ($E_G \sim 1.57$ eV) with a PCE of 15.2 % obtained by successfully substituting the CdS buffer layer with Zn(O,S) buffer layer deposited using chemical bath deposition.[11] By replacing CdS with Zn(O,S) the interface recombination in the device and the difference between qFLs and $V_{OC}$ reduced due to better buffer CBM alignment with CIGSu.[11] We also found that the substitution of CdS with Zn(O,S) buffer layer having higher CBM results in an electrical transport barrier when using ZnO as i-layer in the device. The barrier is apparent as a rollover in the positive power quadrant in the device, which results in low fill factor (FF) and PCE. The conduction band offset between CdS and ZnO has been found to be -0.30±0.10 eV,[8] therefore, the rollover might be caused by a high negative conduction band offset at the Zn(O,S)/ZnO (buffer/i-layer) interface. To achieve high FF in the Zn(O,S) buffer device Al:(Zn,Mg)O i-layer was used, which led to PCE above 15 %.[11] However, the device fabrication process required an additional annealing step (~200 °C for 10 minutes) after buffer deposition. This step leads to a reduction in qFLs of the absorber (see Figure S1 in the supplementary information), therefore limiting the maximum achievable $V_{OC}$ and PCE of the device.

In this contribution, the replacement of the CdS/i-ZnO (buffer/i-layer) stack is explored by a (Zn,Mg)O /Al:(Zn,Mg)O stack using atomic layer deposition (ALD) and magnetron sputtering process that does not require an additional high temperature-annealing step. The impact of varying Mg composition in the (Zn,Mg)O buffer on CIGSu device properties is explored. Calibrated photoluminescence measurements are performed on the absorber to extract the qFLs values. Current-voltage and external quantum efficiency measurements are performed on the completed device to extract the electrical characteristics. The deficit between measured qFLs/e and $V_{OC}$ is investigated and compared for different Mg concentrations in (Zn,Mg)O films. Numerical simulations are performed to comprehend the influence of electrical barriers on device current density-voltage (J-V) characteristics of the fabricated devices. The results presented here show the necessity of properly tuning the conduction band offset at various interfaces in the device to realize high PCE CIGSu devices.

**CIGSu device fabrication**

*Absorber deposition:* Cu-poor CIGSu absorbers were prepared on Mo coated soda-lime glass substrates *via* a 3-stage co-evaporation process at a temperature of ~ 570 ºC with the final film thickness of ~2 μm. The detailed procedure is reported in Ref [11]. The desired absorber stoichiometry was obtained by controlling the evaporation flux of the Cu, In, and Ga sources and the duration of the three stages of the process. The final bulk composition of as-grown absorbers was determined using energy-dispersive x-ray spectroscopy (EDS).

*Buffer deposition:* Prior to buffer deposition, all the absorbers were etched in 5% KCN solution for 30 seconds to remove any oxides, followed by rinsing and storing in de-ionized water. This was done to prevent the possible degradation of optoelectronic properties of chalcopyrite absorbers by air exposure, as observed previously.[16, 17] The samples were blow-dried with nitrogen gas to remove the water layer immediately before introducing them into the ALD reactor. The atomic layer deposition of the (Zn,Mg)O buffer was carried out in an ALD reactor using Diethylzinc (DEZ) and Bis(cyclopentadienyl) magnesium ($Mg(Cpt)_2$) as precursors for ZnO and MgO, respectively, with water as co-reactant. $Mg(Cpt)_2$ was heated at 75-80 °C to achieve high enough vapor pressure. The depositions were carried out between 130-150 °C using a supercycle approach, *i.e.* alternating ZnO cycles ($DEZ/H_2O$) and MgO cycles ($Mg(Cpt)_2/H_2O$) with a ratio chosen in order to give the expected composition. Depending upon the required composition, MgO and ZnO cycles were repeated to achieve a thickness of 30 nm.

*i-layer and transparent conductive oxide widow layer deposition:* Both ALD and magnetron sputtering was used for i-layer deposition. The sputtering process for Al:(Zn,Mg)O is based on previous work on (Zn,Mg)O.[18] The 2-inch targets of MgO with 2 wt % $Al_2O_3$ and of (undoped) ZnO were co-sputtered at a power of 80 W and 110 W, respectively. Throughout the deposition, 1.0 mTorr partial pressure of argon (99.99%) was maintained with the help of a mass flow controller. Desired composition of Al:$Zn_{1-x}Mg_xO$ (x = 0.25) was achieved by tuning the Al:MgO target power.

The composition of deposited films was determined by EDS performed on films grown on Si substrate for ALD deposited (Zn,Mg)O and on molybdenum coated soda-lime glass for sputtered Al:(Zn,Mg)O.

After i-layer deposition, a transparent conductive oxide (TCO) window layer (Al:ZnO) was sputtered onto the samples. For this, a 2 wt % Al-doped ZnO single target was sputtered at 140 W. Other growth conditions were the same as for Al:(Zn,Mg)O deposition. The Ni-Al grids were deposited on the samples using e-beam evaporation to complete the device fabrication process. On each sample, several devices of area ~0.5 $cm^2$ were realized by mechanical scribing.

*Characterization methods:* EDS with an operating voltage of 20 kV was used to determine the stoichiometry of as-grown absorbers and buffer layers. For determining the thickness of the ALD deposited (Zn,Mg)O buffer layer, a Si wafer was placed alongside the CIGSu samples in each run. The thickness of (Zn,Mg)O film on Si wafer was then determined by ellipsometry using an effective medium approximation model. Photoluminescence (PL) measurements were carried out in a home-built system using a steady-state excitation via a CW diode laser of 405 nm wavelength as the excitation source. The raw PL data was acquired through two off-axis mirrors, then spectrally resolved and detected by a monochromator and Si-CCD, respectively. Spectral and intensity corrections are subsequently applied to the raw PL data. The spectral correction is performed with a commercial calibrated halogen lamp. The intensity correction entails the measurement of the laser beam diameter with a charged-couple device (CCD) camera and the laser power by a photodetector. The photon flux from the laser is calculated and adapted to the AM 1.5 solar spectrum photon flux above the bandgap. Thus, an illumination corresponding to '1 sun' for a bandgap of ~1.6 eV is used on the samples. The corrected spectrum is transformed into the energy domain and assessed using Planck's generalized law[19], which describes the dependence of the luminescence yield as a function of absorptivity, temperature, and qFLs. The qFLs is extracted from a fit of the high-energy wing of the PL spectrum where absorptivity is assumed unity and the temperature is fixed to 300 K.[17] A variation in the qFLs value across the absorbers grown in the same run and even on the same absorber was observed (Figure S2). We attribute this difference in qFLs values to the inhomogeneous heating of the sample stage during absorber deposition.

For measuring device J-V properties, a Xenon short-arc lamp AAA-Standard solar simulator was used together with an IV source-measure unit. The simulator was calibrated to air mass 1.5 global using a Si reference cell. For determining the external quantum efficiency (EQE) of devices, EQE setup consisting of halogen and xenon lamps, a grating monochromator, a chopper and calibrated reference diodes was used. The current was measured using a lock-in amplifier and the intensity of the light by calibrated reference diodes. Low-temperature J-V measurements were performed by mounting the samples in a closed-cycle cryostat chamber, where a base pressure below $4\times10^{-3}$ mbar is maintained. For measuring the J-V at one-sun illumination conditions, a cold mirror halogen lamp with an equivalent intensity of 100 mW/cm$^2$ was used. The lamp intensity was adjusted by setting the short-circuit current density ($J_{sc}$) of the sample equal to the value measured under the solar simulator, which was achieved by changing the height of the lamp from the sample. For precise measurement of the sample temperature, a Si-diode sensor glued onto an identical glass substrate was placed just beside the solar cell.

*Numerical simulation*: Numerical simulations were done using the solar cell capacitance simulator (SCAPS-1D), which solves the one-dimensional Poisson and continuity equation for electrons and holes, using appropriate boundary conditions at the interfaces and the contacts. The software is developed at the Department of Electronics and Information Systems (ELIS) of the University of Gent, Belgium, under Marc Burgelman.[20] The device parameters used in this study are reported in table S1.

## Effect of Mg content in the buffer on device properties

We start by exploring the effect of different compositions of $Zn_{1-x}Mg_xO$ buffer layers on the CIGSu device performance. For this purpose, we used ALD deposited $Zn_{1-x}Mg_xO$ films with x = 0.27, 0.30 and 0.37, with a measured bandgap of 3.8 eV, 3.9 eV, and 4.6 eV, respectively. The composition was chosen based on the following argument: the CIGSu/CdS device has an activation energy value ~0.2 eV less than the bulk bandgap of CIGSu for the dominant recombination pathway,[11] which implies that the CdS buffer layer has a CBM that is 0.2 eV lower than the CBM of CIGSu.[21] Next, ZnO is known to have a CBO of -0.30±0.10 eV with CdS[8]. Thus indirectly we can assume CIGSu has a -0.50 eV CBO with ZnO. Now, assuming the increase in bandgap due to Mg incorporation results solely in the increase in CBM of the $Zn_{1-x}Mg_xO$ films, the above composition should yield CIGSu/(Zn,Mg)O devices with CBO of 0.0 eV, 0.1 eV, and 0.8 eV.

The experimental device preparation plan starting from the absorber until TCO deposition is presented in **Figure 1**. The experiments were performed on as grown CIGSu samples having bulk bandgap ~1.6 eV with a [Cu]/[In+Ga] (CGI) ~0.93 and [Ga]/[Ga+In] (GGI) ~0.12. The deficit between qFLs/e (e being the electron charge) and $V_{OC}$ has been established as interface $V_{OC}$ deficit,[22] and is used as the benchmark for making a comparison among different samples. For this, the qFLs is measured at different spots for all the absorbers prior to buffer deposition.

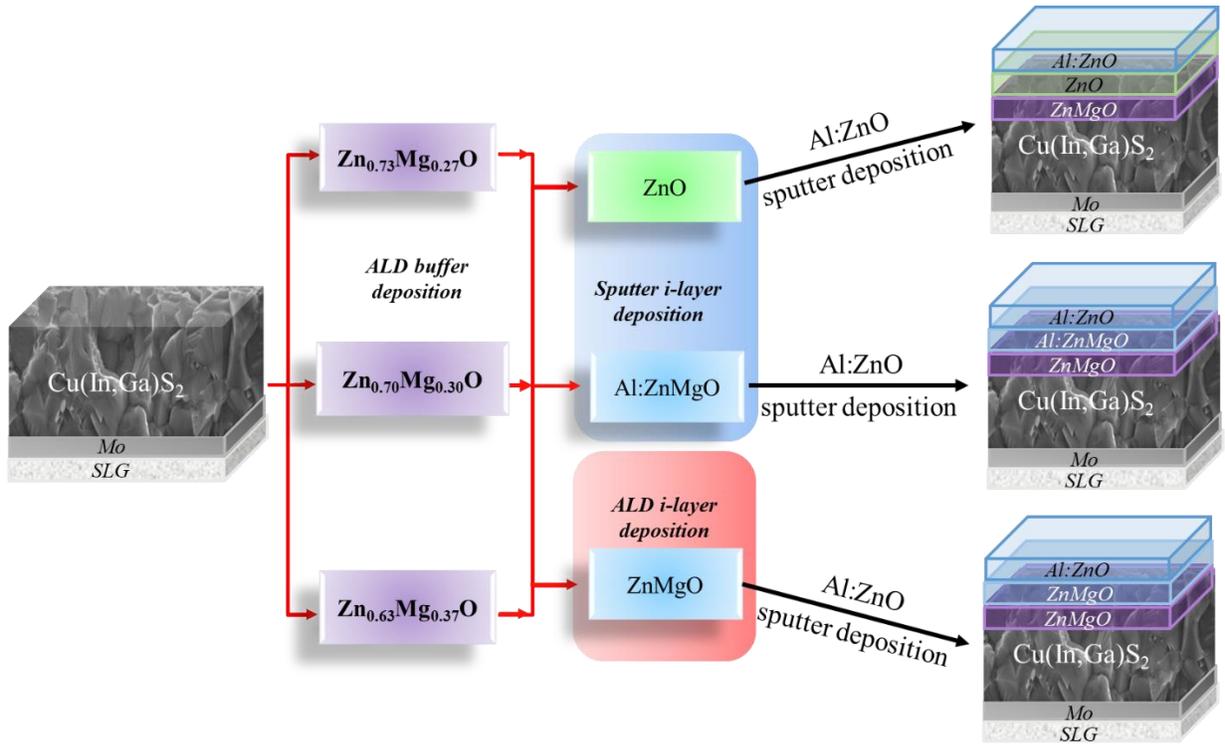

*Figure 1:* *Experimental schematic showing the device fabrication process. The CIGSu absorbers are first processed with $Zn_{1-x}Mg_xO$ buffer layer with x = 0.27, 0.30 and 0.37, followed by deposition of either ZnO or Al:(Zn,Mg)O or (Zn,Mg)O i-layer and finally the deposition of Al:ZnO TCO layer.*

**Figure 2a** shows J-V characteristics of the devices prepared with the three different $Zn_{1-x}Mg_xO$ buffer layers, with either ZnO or Al:(Zn,Mg)O as the i-layer deposited by sputtering. The device J-V characteristic parameters are reported in table 1, along with locally measured qFLs values on the sample before buffer deposition. The device with $Zn_{1-x}Mg_xO$ (x = 0.27) buffer layer and sputtered Al:(Zn,Mg)O i-layer exhibits the highest PCE ~12 % among all devices. Independent of the i-layer used, a deterioration in the device PCE is observed with increasing Mg content in the $Zn_{1-x}Mg_xO$ buffer. Between the two i-layers, lower PCE values are displayed by the devices prepared with ZnO i-layer compared to Al:(Zn,Mg)O i-layer devices with the same buffer composition.

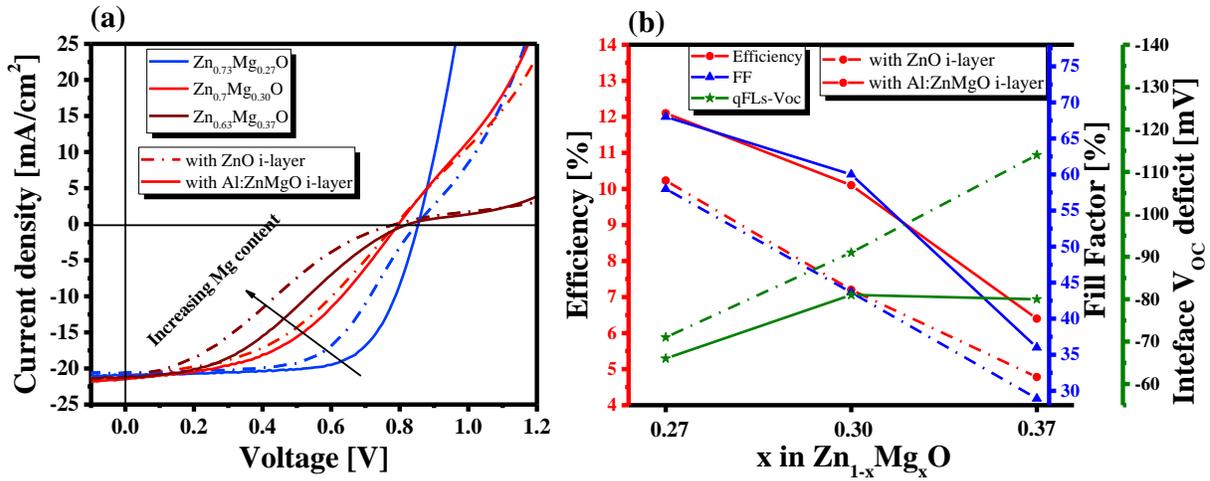

*Figure 2: (a) Measured J-V characteristics of devices with different Mg content in (Zn,Mg)O buffer layers. (b) Scatter chart of device PCE (left axis), fill factor (right axis), and interface $V_{OC}$ deficit (second right axis) vs. different Mg content in (Zn,Mg)O buffer layers. The solid lines represent the device with an i-Al:(Zn,Mg)O+AZO layer and the dash-dotted lines represent the device with i-ZnO+AZO layer on top of the buffer.*

*Table 1: J-V characteristic of best CIGSu devices made of $Zn_{1-x}Mg_xO$ buffer with x~ 0.27, 0.30 and 0.37 and with a sputter-deposited i-ZnO layer or Al:(Zn,Mg)O i-layer and AZO TCO layer. The qLFs values reported are for absorbers before buffer deposition.*

| Zn$_{1-x}$Mg$_x$O | PCE (%) | FF (%) | J$_{sc}$ (mA/cm$^2$) | V$_{OC}$ (mV) | qFLs @1sun (meV) | V$_{OC}$- qFLs/e (mV) |
|---|---|---|---|---|---|---|
| x=0.27 Al:ZnMgO i-layer | 12.1 | 68 | 20.9 | 854 | 920 | -68 |
| x=0.27 ZnO i-layer | 10.1 | 58 | 20.6 | 850 | 921 | -71 |
| x=0.3 Al:ZnMgO i-layer | 10.1 | 60 | 21.2 | 796 | 877 | -81 |
| x=0.3 ZnO i-layer | 7.2 | 44 | 20.9 | 785 | 876 | -91 |
| x=0.37 Al:ZnMgO i-layer | 6.4 | 36 | 21.5 | 825 | 905 | -80 |
| x=0.37 ZnO i-layer | 4.8 | 29 | 21.0 | 787 | 901 | -114 |

**Figure 2b** shows the variation of PCE, FF, and the interface V$_{OC}$ deficit of the devices as a function of the Mg content for devices prepared with both ZnO and Al:(Zn,Mg)O i-layer. The drop in PCE with increasing Mg content in buffer layers can be majorly attributed to a loss of FF (**Figure 2b**). In the case of ZnO i-layer devices, the loss in FF is even higher than the Al:(Zn,Mg)O i-layer devices therefore the PCE is lower. The loss in FF is due to distortion in the J-V curve of the devices in the form of 'S shape' in the negative power quadrant and rollover in the positive power quadrant, as evident in the J-V curve of devices (**Figure 2a**).

With the increasing Mg content in the buffer, the interface V$_{OC}$ deficit increases, particularly in the devices prepared with the ZnO i-layer. This indicates an increased qFLs loss in the device. The loss originates partially from the absorber degradation after buffer deposition, as the qFLs measurements display a drop in qFLs after (Zn,Mg)O buffer deposition (see Figure S4, performed on a different sample). Further loss in V$_{OC}$ relative to qFLs measured after buffer deposition might originate from the gradient in electron qFL either in the vicinity of the absorber surface[22] or within the contact layer. Similar losses have been observed in selenide solar cells with non-optimized contact layers.[17] The presence of negative conduction band offset (CBO) at the CIGSu/(Zn,Mg)O interface could be one possible origin of this loss. However, the temperature-dependent $V_{OC}$ measurements on the device with the lowest CBM in the $Zn_{1-x}Mg_xO$ buffer, *i.e.* the one with 'x' = 0.27 yield activation energy of the dominant recombination path equal to the bulk bandgap (see Figure S3). This excludes the presence of a negative CBO CIGSu/(Zn,Mg)O interface, otherwise, the activation energy would have been less than the bulk bandgap.[21, 23] We show later by simulations that the CBOs at the buffer/i-layer interface are critical parameters for the $V_{OC}$ loss.

The above results suggest that the $Zn_{1-x}Mg_xO$ (x = 0.27) buffer layer with sputtered Al:(Zn,Mg)O i-layer is the best combination for high device PCE. With the optimized buffer conditions, another CIGSu device with ALD deposited $Zn_{1-x}Mg_xO$ (x = 0.27) buffer layer and sputtered Al:(Zn,Mg)O i-layer was fabricated and equipped with an anti-reflective coating. The absorber used for this device possessed a qFLs of 986 meV with a $E_G$ ~1.63 eV obtained by $\frac{d(EQE)}{dE}$ analysis (see Figure S5). The higher $E_G$ is due to slightly higher Ga concentration in the absorber [Ga]/[Ga+In]~0.18 compared to earlier devices. The J-V characteristic of the device is presented in **Figure 3.** The device exhibits a PCE of 14.0 % with a $V_{OC}$ of 943 mV after light soaking of 30 minutes under open-circuit conditions. The $V_{OC}$ obtained is the highest among all the devices fabricated in this study. Although even this device possesses a moderate interface $V_{OC}$ deficit of ~ 40 mV. The higher PCE is a consequence of the improved optoelectronic quality of the absorber and interface quality of the device. The device has a 30 mV lower qFLs/e loss w.r.t. to Shockley–Queisser $V_{OC}$[24] and 28 mV lower interface $V_{OC}$ deficit compared to the best devices discussed previously. The results thus show that a combination of $Zn_{1-x}Mg_xO$ (x = 0.27) buffer layer and sputtered Al:(Zn,Mg)O is ideal for CIGSu devices with a bandgap around 1.6 eV. Achieving even higher PCE needs further improvement in the optoelectronic properties of the absorber.

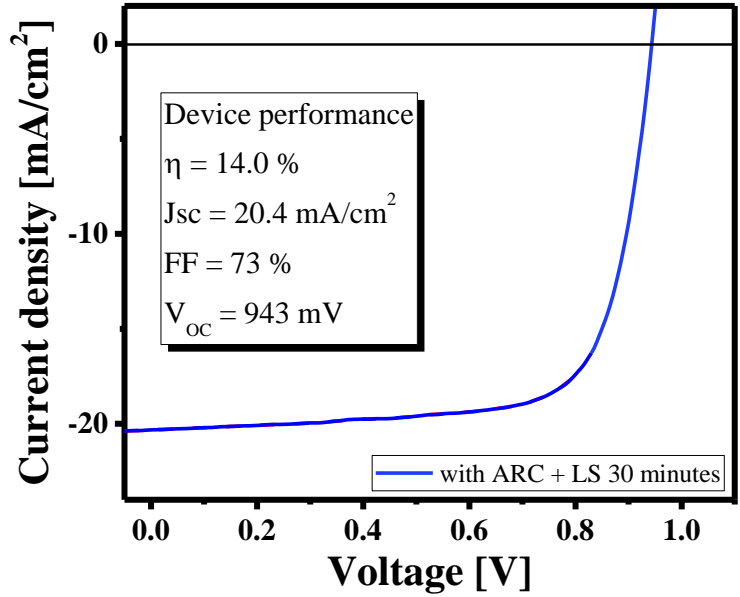

*Figure 3: J-V curve of the best device prepared with $Zn_{1-x}Mg_xO$ (x = 0.27) buffer layer together with Al:(Zn,Mg)O i-layer and AZO TCO layer. The curve is obtained for a device with an anti-reflective coating of $MgF_2$ and light-soaked for 30 minutes under open-circuit conditions.*

## Numerical simulations for electrical barriers

The J-V curves of the devices with x ≥ 0.3 in the $Zn_{1-x}Mg_xO$ buffer layer exhibit 'S shape' in the negative power quadrant and a rollover at a forward bias above their $V_{OC}$ (**Figure 2**). This is a typical sign of an extraction and injection barrier at the front or back contact for charge carriers[25, 26]. The back contact barrier can be excluded as the device with $Zn_{1-x}Mg_xO$ (x = 0.27) buffer layer doesn't display distortion in the J-V curve. Therefore, the front contact barrier is the only viable cause for the 'S shape' and rollover in these devices[26]. This, of course, assumes that ALD buffer deposition does not change absorber properties near the back contact nor the properties of the back contact itself. Using numerical simulations, the effect of a front barrier on the device J-V characteristics will be explored here.

The electrical barrier in the devices studied here likely originates from the increased CBM of the (Zn,Mg)O buffer films with increased Mg content, as observed earlier by Minemoto et al.[14] The increased CBM also increases the CBO at the absorber/buffer and the buffer/i-layer interface, thus leading to an electrical barrier for photogenerated electrons and the injected electrons. **Figure 4a** shows a simulated conduction band profile of the junction, depicting the two barriers: one for the photogenerated carriers $\varphi_b^{ph}$, other for the injected electrons $\varphi_b^{in}$. $\varphi_b^{ph}$ is the energetic distance between the electron Fermi-level at the absorber/buffer interface and the CBM of the buffer at the absorber/buffer interface, and $\varphi_b^{in}$ is the energetic distance between the electron Fermi-level at the i-layer/TCO interface and the CBM of the buffer at the absorber/buffer interface.[21] The $\Delta E_{Fn}$ is the drop in Fermi-level at the interface, which is required to derive the diode current across the junction.

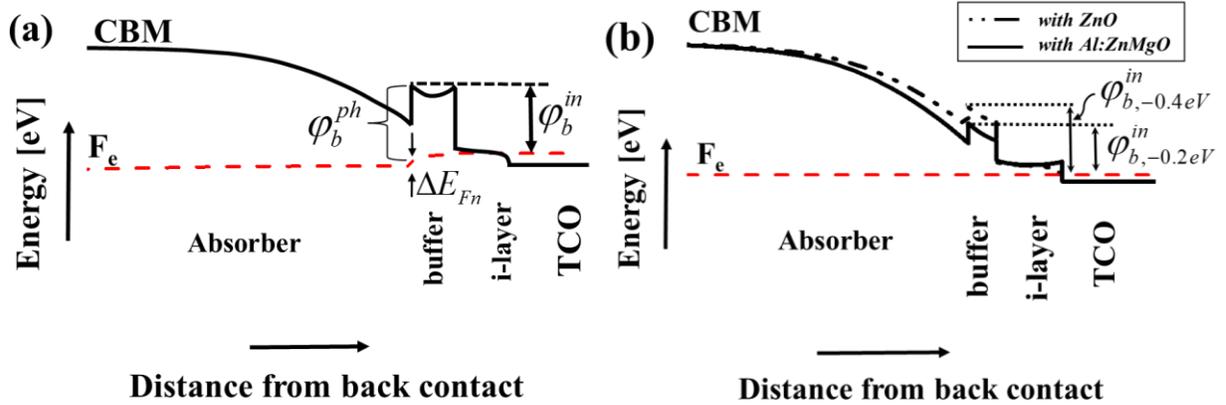

***Figure 4:*** *(a) Simulated band diagram of a device showing the injection $\varphi_b^{in}$ and extraction barrier $\varphi_b^{ph}$ in a device at forward bias. (b) Simulated equilibrium energy band diagram of devices one with the Al:(Zn,Mg)O i-layer (solid) and one with ZnO i-layer (dash-dotted lines). For simulating the Al:(Zn,Mg)O i-layer device a CBO of -0.2 eV is introduced at both buffer/i-layer interface and i-layer/TCO interface, whereas for simulating the ZnO i-layer device a CBO of -0.4 eV is introduced at the buffer/ZnO i-layer interface, and a flat conduction band at i-layer/TCO interface.*

From **Figure 4a**, it is clear that the height of the two barriers is equal at equilibrium and is dependent on the position of electron Fermi-level at the absorber/buffer interface and position of electron Fermi-level at the i-layer/TCO interface, for $\varphi_b^{ph}$ and $\varphi_b^{in}$ respectively. While it is evident that the two barriers are dependent on the CBM of the buffer, the $\varphi_b^{in}$ is also impacted by the CBM of i-layer as well. This is because it changes the band bending in the absorber and, therefore, the position of electron Fermi-level at the i-layer/TCO interface. For example, **Figure 4b** depicts simulated band diagrams of two devices:

1. Device with a CBO of 0.2 eV at the absorber/buffer interface, -0.4 eV at the buffer/i-layer interface, and no CBO at the i-layer/TCO interface (the dash-dotted line **Figure 4b**).

2. Device with a CBO of 0.2 eV at the absorber/buffer interface, -0.2 eV at both buffer/i-layer and i-layer/TCO interfaces (the solid line **Figure 4b**).

The CBM of the i-layer in the second device is 0.2 eV higher than the first device (based on observed 0.2 eV increase in the bandgap for Al:(Zn,Mg)O compared to ZnO see Figure S6). Here we have tried to simulate a situation comparable to the case where ZnO or Al:(Zn,Mg)O assumed is used as i-layer in the device. Even though CBO at the absorber/buffer interface and CBO between buffer and the TCO layer is the same, the barrier height in equilibrium $\varphi_b^{in}$ is different in each case for the two devices. This is because the band bending in the absorber depends on the buffer-i-layer CBO. For device 1 it is 0.510 eV, whereas for device 2 it is 0.678 eV. This shows that the CBM of the i-layer also has an influence on the barrier height in the device.

By varying Mg concentration in buffer and i-layer, we vary either the CBM of the buffer or the CBM of the i-layer in the devices prepared in this study above. An increase in $\varphi_b^{ph}$ and $\varphi_b^{in}$ is expected with the increase in Mg concentration in the $Zn_{1-x}Mg_xO$ buffer layer and a decreases with an increase in the i-layer, as it is known to increase the CBM of the films.[14] A qualitative method to differentiate and comprehend the impact of different Mg contents in buffer and i-layer on the electrical barriers and the device J-V characteristics is to simulate the device in the SCAPS 1-D simulator.[20, 27] In the following, the effect of these two barriers is explored with the help of the SCAPS 1-D simulator, first separately and then together on device J-V characteristics.

Below we discuss three scenarios: (i) varying CBO at the absorber/buffer interface with constant CBO at the buffer/i-layer, (ii) varying CBO at the buffer/i-layer interface with constant CBO at the absorber/buffer interface, and (iii) finally varying the CBM in the buffer, thus varying CBO at the absorber/buffer and the buffer/i-layer interface. The first two cases particularly help to distinguish the effect of the CBO at the absorber/buffer interface and the buffer/i-layer interface on the J-V curves of the device separately. The parameters used for these simulations are reported in table S1.

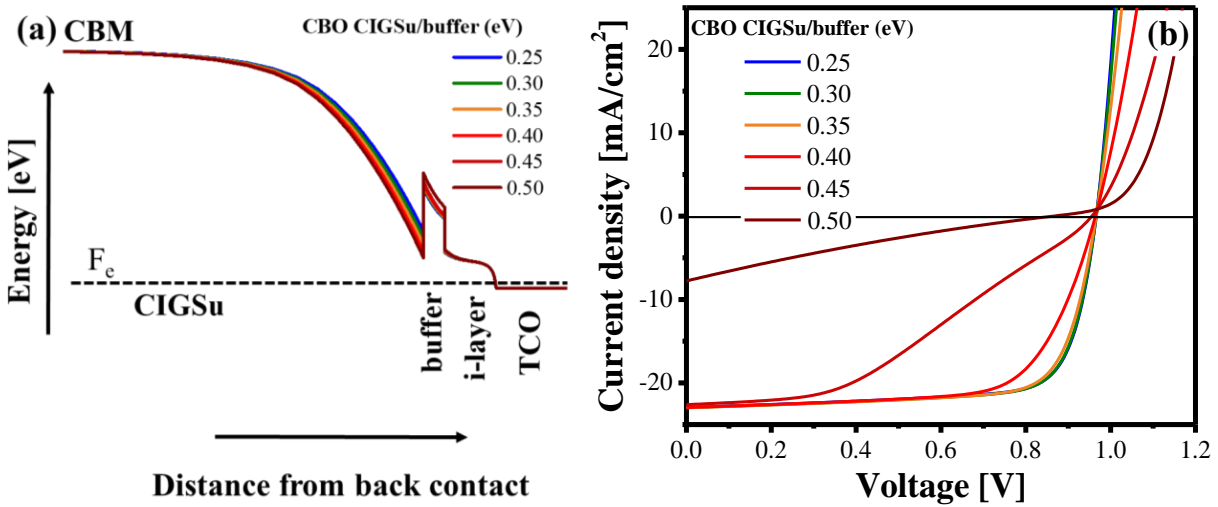

*Figure 5:* (a) Simulated band diagram and (b) J-V curves of CIGSu device with varying CBO at CIGSu/buffer interface maintaining a constant CBO (-0.2 eV) at buffer/i-layer interface and no CBO at the i-layer/TCO layer.

We start by looking at the impact of scenario (i) on device J-V curves. While there is no experimental equivalent of this situation in this study, it helps us differentiate the effect of CBO at the absorber/buffer from the CBO at the buffer/i-layer (TCO layer) interface. For this, a CIGSu device is simulated in SCAPS-1D, the buffer CBM energy is varied *via* changing its electron affinity. A CBO of -0.2 eV at the buffer/i-layer interface and a flat CBO at the i-layer/TCO interface was maintained by changing the CBM of i-layer and TCO equal to the change in buffer CBM energy. This was done to exclude the effects of change in CBO at the buffer/i-layer interface and i-layer TCO on the J-V curves of the device. The CBO at the CIGSu/buffer interface was varied from 0.25 eV to 0.50 eV.

The equilibrium band diagram and the J-V characteristics are simulated and presented in **Figure 5**, respectively. It can be observed that CBO value ≤ 0.35 eV at the CIGSu/buffer interface does not impact the J-V curves significantly. However, a CBO value above 0.35 eV results in an 'S shape' in the negative power quadrant, reducing the FF of the device. A CBO ≥ 0.50 eV starts affecting the $J_{SC}$ and $V_{OC}$ as well. The 'S shape' originates from a high $\varphi_b^{ph}$ for CBO > 0.4 eV as evident in **Figure 5a**, which limits the flow of photogenerated electrons from the absorber to the buffer and thus photocurrent in the device. Consequently, a low FF in case of moderate barrier height and $J_{SC}$ and $V_{OC}$ in the case of high barrier height is observed.

It must be noted that although such a device with high $\varphi_b^{ph}$ as evident from **Figure 5b** exhibits 'S shape' and consequently a very low FF, it does not exhibit rollover in the positive power quadrant as observed in **Figure 2**.

Next, let us look at the impact of scenario (ii) on the device J-V characteristics with no CBO at the CIGSu/buffer interface. The CIGSu device is simulated with different CBO at the buffer/i-layer interface, maintaining flat CBO at the CIGSu/buffer interface. The barrier $\varphi_b^{in}$ was varied by varying the CBM of the i-layer and keeping the CBM of the buffer layer fixed at a value equal to the CBM of the absorber (*i.e.* with equal electron affinity). Such a device can be envisioned experimentally by varying Mg content in the (Zn,Mg)O i-layer without altering the Mg content in the $Zn_{1-x}Mg_xO$ buffer layer.

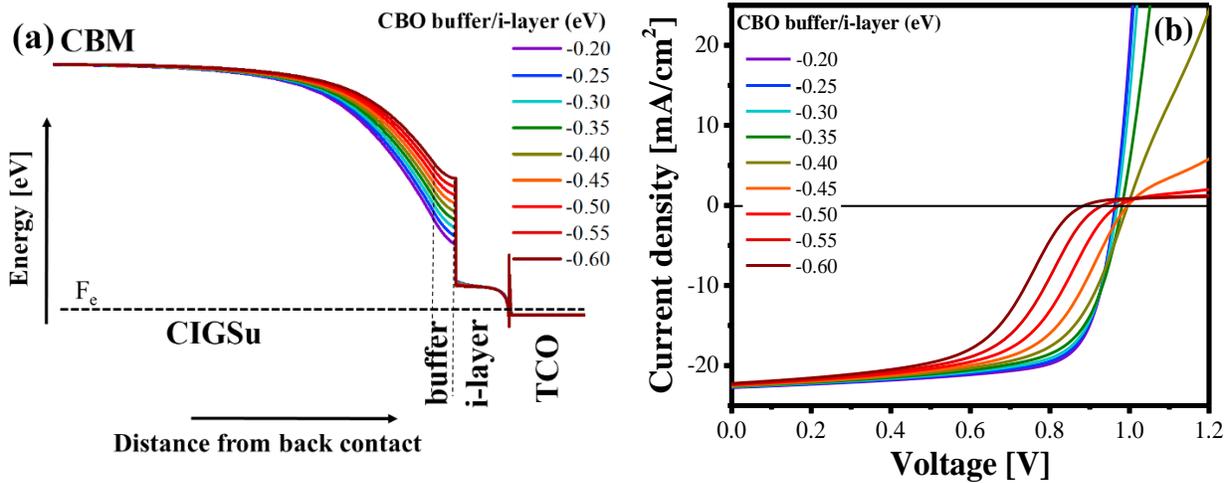

*Figure 6: (a) Simulated band diagram and (b) J-V curves of CIGSu device with varying CBO at buffer/i-layer interface by changing the CBM of the i-layer and keeping the CBM of CIGSu and buffer layer at a fixed value to get a flat CBO at CIGSu/buffer interface.*

**Figure 6a and b** show the simulated band diagram and the J-V characteristics of such a device. The FF of the simulated devices deteriorates as the CBO at the buffer/i-layer interface decreases below -0.25 eV. Moreover, even the $V_{OC}$ starts deteriorating for CBO values below < -0.50 eV. The drop in FF is a direct consequence of an 'S shape' in the form of a strong rollover in the positive power quadrant, which suggests blocking on injected carriers or conventional diode current in forward bias. The rollover is observed because the $\varphi_b^{in}$ increases with increasing negative CBO at the buffer/i-layer interface (see **Figure 6a**). Consequently, a higher potential drop is required in between the TCO and the buffer layer to drive the same amount of diode current through the device as for the lower negative CBO (as observed in **Figure 6b**). The $\varphi_b^{in}$ thus acts as a series resistance in the device and thus leads to a drop in FF. As the CBO or the $\varphi_b^{in}$ at the buffer/i-layer increases the series resistance also increases and the FF decreases.

It must be noted that when compared to the device with a high CBO at the absorber/buffer interface, *i.e.* case (i), the drop in FF is not that substantial in these devices (see **Figure 6b**). This fact is crucial as it helps distinguish the effect of two barriers on the J-V properties of a device.

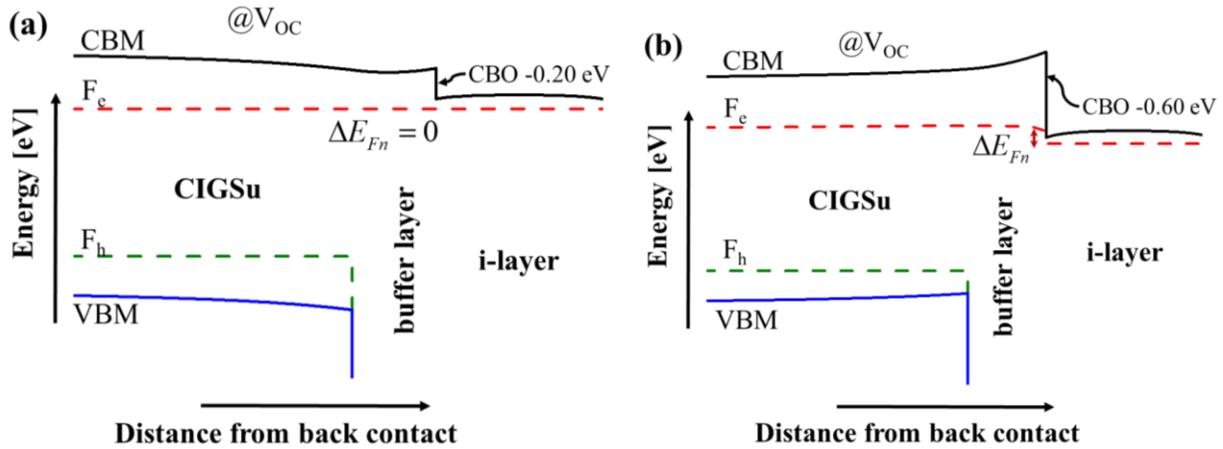

*Figure 7: (a) and (b) shows the simulated band diagram of the CIGSu device at open-circuit conditions with -0.20 eV CBO and -0.60 eV CBO at the buffer/i-layer interface, respectively.*

To understand the origin of $V_{OC}$ reduction, the band diagram of two exemplary devices are simulated at open-circuit conditions: one with a low negative CBO -0.2 eV and a high negative -0.6 eV CBO at the buffer/i-layer interface. **Figure 7a and b** show the respective simulated band diagram. The drop in $V_{OC}$ originates from the drop in electron Fermi-level, as a drop in electron Fermi-level near the buffer/i-layer interface is observed for the device with high negative CBO, something that is absent in the low negative CBO device. This explains why the devices with ZnO i-layer have a higher interface $V_{OC}$ deficit compared to Al:(Zn,Mg)O i-layer devices (Table 1).

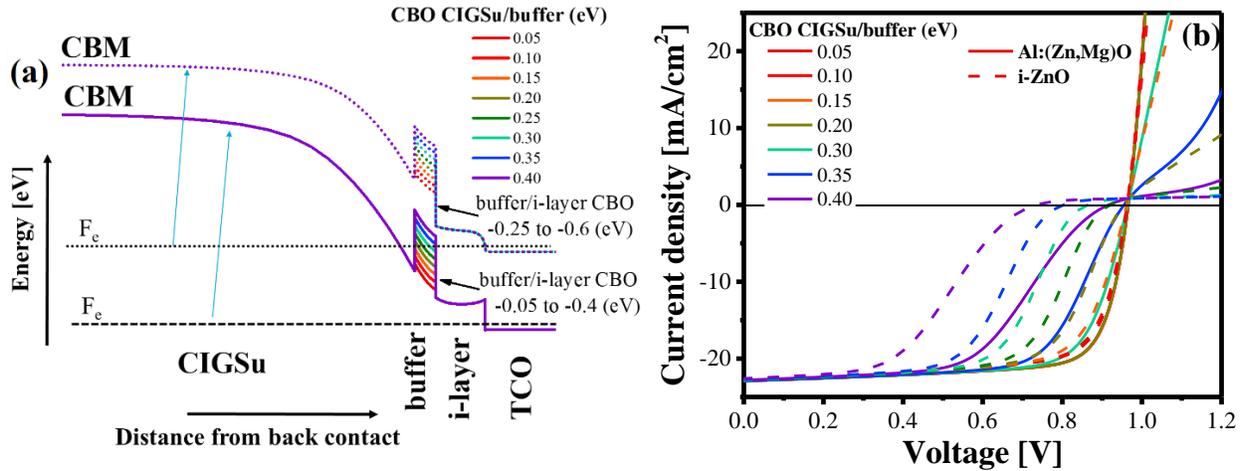

*Figure 8: Simulated band diagram off-setted for clarity (a) and J-V curves of CIGSu device (b) with varying CBM position of buffer with either ZnO i-layer (dotted lines) or Al:(Zn,Mg)O i-layer (solid lines). With ZnO the CBO at buffer/i-layer interface varies between -0.25 eV to -0.60 eV, whereas for Al:(Zn,Mg)O, it varies between -0.05 eV to -0.40 eV.*

Finally, **Figure 8a and b** show the simulated band diagram and the J-V characteristics of the CIGSu device with varying CBO at the CIGSu/buffer interface and buffer/i-layer interface. Experimentally such a situation could be achieved by varying the Mg content in the $Zn_{1-x}Mg_xO$ buffer layer and keeping the Mg content fixed in the i-layer. Such a scenario most accurately fits the experimental part of this study. Assuming the increased bandgap in Al:(Zn,Mg)O i-layer is due to increased CBM energy, two device structures are simulated: one where the CBO at the CIGSu/buffer was varied from 0.05 eV to 0.40 eV, which consequently results in a -0.25 eV to -0.60 eV CBO at the buffer/i-layer, where i-layer is assumed to be Al:(Zn,Mg)O. Another, where the CBO at the CIGSu/buffer was also varied from 0.05 eV to 0.40 eV, but resulting in a -0.45 eV to -0.80 eV CBO at the buffer/i-layer, where i-layer is assumed to be ZnO. This was achieved by varying just the CBM of the buffer layer and keeping the CBM of the i-layer fixed at a value (4.4 eV and 4.6 eV, respectively for Al:$Zn_{1-x}Mg_xO$ and ZnO i-layers).

The simulated J-V curves display 'S shape' in the negative power quadrant and rollover in the positive power quadrant as the CBM of buffer and consequently the CBO at the CIGSu/buffer interface increases beyond a specific value, which is different for both structures. For the Al:(Zn,Mg)O i-layer device, a significant distortion in the J-V curve is observable for CBO values above 0.35 eV, whereas, for the ZnO i-layer device, the S-shape and rollover are observable even for CBO values above 0.20 eV (see **Figure 8b**). Consequently, in both cases, a drop in FF and $V_{OC}$ is observed. Interestingly, for the same CBO value at the absorber/buffer interface, the drop in FF and $V_{OC}$ is higher for ZnO i-layer simulated device. This is due to a higher negative CBO at the buffer/i-layer interface, as already discussed in case (ii). It is also in agreement with the experimental observation made in **Figure 2**. Moreover, the drop in FF and $V_{OC}$ increases as the CBM of the buffer increases in the simulations, similar to the observations made in **Figure 2**, where increasing Mg content in the (Zn,Mg)O buffer resulted in lower FF and $V_{OC}$ independent of the i-layer. Therefore, the electrical barriers in the device can explain successfully the 'S shape' and rollover and, consequently, the loss in FF and $V_{OC}$ observed in experiments.

To summarize, a high positive CBO at the absorber/buffer interface limits the photocurrent due to a high $\varphi_b^{ph}$, thus leading to 'S shape' in the negative power quadrant. A high negative CBO at the buffer/i-layer on the other hand reduces the diode current in the device due to a higher $\varphi_b^{in}$. Consequently, this leads to an increased series resistance in the device, and in some cases to a 'S shape' in the positive power quadrant. In both scenarios, the FF of the device decreases due to electrical barriers.

## Conclusions

Atomic layer deposited (Zn,Mg)O and sputtered Al:(Zn,Mg)O were explored as a replacement for CdS buffer layer and ZnO i-layer in CIGSu solar cells with $E_G \sim 1.6$ eV. An optimum Mg concentration 'x' = 0.27 in the $Zn_{1-x}Mg_xO$ buffer layer together with 'x' = 0.25 Al:$Zn_{1-x}Mg_xO$ i-layer is found to deliver high device 14 % PCE in the CIGSu device with bandgap $\sim 1.6$ eV.

Mg content above x = 0.30 in the $Zn_{1-x}Mg_xO$ buffer leads to an 'S shape' in the negative power quadrant because of photocurrent blocking due to a high CBO at the absorber/buffer interface. In addition, it also leads to roll over in the positive power quadrant, which increases with the increasing Mg content in the $Zn_{1-x}Mg_xO$ buffer because the negative CBO at the $Zn_{1-x}Mg_xO$/i-layer interface increases. The Al:(Zn,Mg)O i-layer outperforms the traditional ZnO i-layer in terms of PCE as it causes improvement in FF and the interface $V_{OC}$ deficit. The gain is associated with a higher CBM and, therefore, lower CBO at the buffer/i-layer interface. Thus we confirm Al:(Zn,Mg)O as a good substitute for traditional ZnO i-layer in CIGSu solar cells.

Numerical simulations show that a high conduction band offset at the absorber/buffer interface alone leads to photocurrent blocking and therefore 'S shape' in the negative power quadrant. Also, a high conduction band offset at the buffer/i-layer interface leads to diode current blocking and, therefore, a rollover in the positive power quadrant. When present together, they lead to 'S shape' in the negative power quadrant and rollover in the positive power quadrant and, consequently, reduce the FF of the device. In addition, a high negative CBO at the buffer/i-layer interface results in a drop in electron Fermi-level at the interface, causing interface $V_{OC}$ deficit in the devices.

In general, not only the CBO at the absorber/buffer but also the buffer/i-layer interface has a significant impact on solar cell characteristics. For the CIGSu device, the alternative buffer layers and i-layers must be chosen such that they have minimum CBO at the absorber/buffer and buffer/i-layer interface. Also, a low-temperature deposition process must be chosen for buffer and i-layer to preserve the qFLs and enhance the $V_{OC}$ and, consequently, the power conversion efficiency of the device.

## Acknowledgments

This research was funded in whole, or in part, by the Luxembourg National Research Fund (FNR), grant reference [PRIDE 15/10935404/MASSENA] and [PRIDE17/12246511/PACE] project. For the purpose of open access, the author has applied a Creative Commons Attributions 4.0 International (CC BY 4.0) license to any Author Accepted Manuscript version arising from this submission.

We are also thankful to Dr. Marc Burgelman and his team at the University of Ghent, Belgium, for providing SCAPS-1D simulation software.

## References

[1] P. Jackson *et al.*, "Properties of Cu (In, Ga) Se2 solar cells with new record efficiencies up to 21.7%," *Phys. Status. Solidi. R.,* vol. 9, no. 1, pp. 28-31, 2015, doi: https://doi.org/10.1002/pssr.201409520.
[2] P. Jackson *et al.*, "New world record efficiency for Cu (In, Ga) Se2 thin-film solar cells beyond 20%," *Prog. Photovolt.,* vol. 19, no. 7, pp. 894-897, 2011, doi: https://doi.org/10.1002/pip.1078.
[3] P. Jackson, R. Wuerz, D. Hariskos, E. Lotter, W. Witte, and M. Powalla, "Effects of heavy alkali elements in Cu(In,Ga)Se2 solar cells with efficiencies up to 22.6%," *Phys. Status Solidi RRL,* vol. 10, no. 8, pp. 583-586, 2016, doi: 10.1002/pssr.201600199.
[4] R. Carron *et al.*, "Advanced Alkali Treatments for High-Efficiency Cu (In, Ga) Se2 Solar Cells on Flexible Substrates," *Adv. Mater.,* vol. 9, no. 24, p. 1900408, 2019, doi: https://doi.org/10.1002/aenm.201900408.
[5] T. Kato, J.-L. Wu, Y. Hirai, H. Sugimoto, and V. Bermudez, "Record efficiency for thin-film polycrystalline solar cells up to 22.9% achieved by Cs-treated Cu (In,Ga)(Se,S)2," *IEEE J. Photovolt.,* vol. 9, no. 1, pp. 325-330, 2018, doi: https://doi.org/10.1109/JPHOTOV.2018.2882206.
[6] A. Miguel *et al.*, "Diode characteristics in state-of-the-art ZnO/CdS/Cu (In1-x Gax) Se2 solar cells," *Prog. Photovolt.: Res. Appl,* vol. 13, pp. 209-216, 2005.

[7] L. Weinhardt, C. Heske, E. Umbach, T. Niesen, S. Visbeck, and F. Karg, "Band alignment at the i-ZnO/CdS interface in Cu (In, Ga)(S, Se) 2 thin-film solar cells," *Appl. Phys. Lett.,* vol. 84, no. 16, pp. 3175-3177, 2004, doi: https://doi.org/10.1063/1.1704877.

[8] G. V. Rao, F. Säuberlich, and A. Klein, "Influence of Mg content on the band alignment at Cd S∕(Zn, Mg) O interfaces," *Appl. Phys. Lett.,* vol. 87, no. 3, p. 032101, 2005, doi: https://doi.org/10.1063/1.1995951.

[9] B. Tell, J. Shay, and H. J. P. r. B. Kasper, "Electrical Properties, Optical Properties, and Band Structure of CuGaS2 and CuInS2," *Phys. Rev. B.,* vol. 4, no. 8, p. 2463, 1971, doi: https://doi.org/10.1103/PhysRevB.4.2463.

[10] L. Weinhardt, O. Fuchs, D. Groß, G. Storch, E. Umbach, N. G. Dhere, A. A. Kadam, S. S. Kulkarni, and C. Heske, "Band alignment at the CdS∕Cu(In, Ga)S2 interface in thin-film solar cells," *Appl. Phys. Lett.,* vol. 86, no. 6, p. 062109, 2005, doi: https://doi.org/10.1063/1.1861958.

[11] S. Shukla *et al.*, "Over 15% efficient wide-band-gap Cu (In, Ga) S2 solar cell: Suppressing bulk and interface recombination through composition engineering," *Joule,* 2021, doi: https://doi.org/10.1016/j.joule.2021.05.004.

[12] C. Persson, C. Platzer-Björkman, J. Malmström, T. Törndahl, and M. Edoff, "Strong valence-band offset bowing of ZnO 1− x S x enhances p-type nitrogen doping of ZnO-like alloys," *Phys. Rev. Lett.,* vol. 97, no. 14, p. 146403, 2006, doi: https://doi.org/10.1103/PhysRevLett.97.146403.

[13] D. Kieven *et al.*, "Band alignment at sputtered ZnSx O1–x/Cu (In, Ga)(Se, S) 2 heterojunctions," *Phys. Status. Solidi. R.,* vol. 6, no. 7, pp. 294-296, 2012, doi: https://doi.org/10.1002/pssr.201206195.

[14] T. Minemoto, Y. Hashimoto, T. Satoh, T. Negami, H. Takakura, and Y. Hamakawa, "Cu (In, Ga) Se 2 solar cells with controlled conduction band offset of window/Cu (In, Ga) Se 2 layers," *J. Appl. Phys.,* vol. 89, no. 12, pp. 8327-8330, 2001, doi: https://doi.org/10.1063/1.1366655.

[15] H. Hiroi, Y. Iwata, S. Adachi, H. Sugimoto, and A. Yamada, "New World-Record Efficiency for Pure-Sulfide Cu(In,Ga)S$_2$ Thin-Film Solar Cell With Cd-Free Buffer Layer via KCN-Free Process," *IEEE J. Photovolt.,* vol. 6, no. 3, pp. 760-763, 2016, doi: https://doi.org/10.1109/JPHOTOV.2016.2537540.

[16] D. Regesch *et al.*, "Degradation and passivation of CuInSe 2," *Appl. Phys. Lett.,* vol. 101, no. 11, p. 112108, 2012, doi: https://doi.org/10.1063/1.4752165.

[17] F. Babbe, L. Choubrac, and S. Siebentritt, "Quasi Fermi level splitting of Cu-rich and Cu-poor Cu (In, Ga) Se2 absorber layers," *Appl. Phys. Lett.,* vol. 109, no. 8, p. 082105, 2016, doi: https://doi.org/10.1063/1.4961530.

[18] Y. Inoue, M. Hála, A. Steigert, R. Klenk, and S. Siebentritt, "Optimization of buffer layer/i-layer band alignment," in *2015 IEEE 42nd Photovoltaic Specialist Conference (PVSC)*, 2015: IEEE, pp. 1-5, doi: https://doi.org/10.1109/PVSC.2015.7355902.

[19] P. Wurfel, "The chemical potential of radiation," *Journal of Physics C: Solid State Physics,* vol. 15, no. 18, p. 3967, 1982, doi: https://doi.org/10.1088/0022-3719/15/18/012.

[20] M. Burgelman, P. Nollet, and S. Degrave, "Modelling polycrystalline semiconductor solar cells," *Thin Solid Films,* vol. 361-362, pp. 527-532, 2000/02/21/ 2000, doi: https://doi.org/10.1016/S0040-6090(99)00825-1.

[21] R. Scheer and H. Schock, "Thin film heterostructures," *Chalcogenide Photovoltaics,* pp. 9-127, 2011, doi: https://doi.org/10.1002/9783527633708.ch2.

[22] M. Sood *et al.*, "Near surface defects: Cause of deficit between internal and external open-circuit voltage in solar cells," *Prog. Photovolt.,* vol. 30, no. 3, pp. 263-275, 2021, doi: 10.1002/pip.3483.

[23] R. Scheer, "Activation energy of heterojunction diode currents in the limit of interface recombination," *J. Appl. Phys.,* vol. 105, no. 10, p. 104505, 2009, doi: https://doi.org/10.1063/1.3126523.

[24] W. Shockley and H. J. Queisser, "Detailed balance limit of efficiency of p-n junction solar cells," *J. Appl. Phys.,* vol. 32, no. 3, pp. 510-519, 1961, doi: https://doi.org/10.1063/1.1736034.

[25] A. Niemegeers and M. Burgelman, "Effects of the Au/CdTe back contact on IV and CV characteristics of Au/CdTe/CdS/TCO solar cells," *J. Appl. Phys.,* vol. 81, no. 6, pp. 2881-2886, 1997, doi: https://doi.org/10.1063/1.363946.

[26] R. Scheer and H. Schock, "Appendix A: Frequently Observed Anomalies," *Chalcogenide Photovoltaics,* pp. 305-314, 2011, doi: https://doi.org/10.1002/9783527633708.ch7.

[27] A. Niemegeers, S. Gillis, and M. Burgelman, "A user program for realistic simulation of polycrystalline heterojunction solar cells: SCAPS-1D," in *Proceedings of the 2nd World Conference on Photovoltaic Energy Conversion, JRC, European Commission, juli*, 1998, pp. 672-675, doi: https://biblio.ugent.be/publication/281948.

**SI: Electrical barriers and their elimination by fine-tuning ZnMgO composition in Cu(In,Ga)S$_2$: Systematic approach to achieve over 14% power conversion efficiency**

Mohit Sood*[1], Poorani Gnanasambandan[1,2], Damilola Adeleye[1], Sudhanshu Shukla[1], Noureddine Adjeroud[2], Renaud Leturcq[2], Susanne Siebentritt[1]

(E-mail address: mohit.sood@uni.lu)

[1]*Department of Physics and Materials Science, University of Luxembourg, Belvaux, L-4422, Luxembourg*

[2]*Material Research and Technology department, Luxembourg Institute of Science and Technology, Belvaux, L-4422, Luxembourg*

**Influence of annealing during or after buffer deposition on qFLs:**

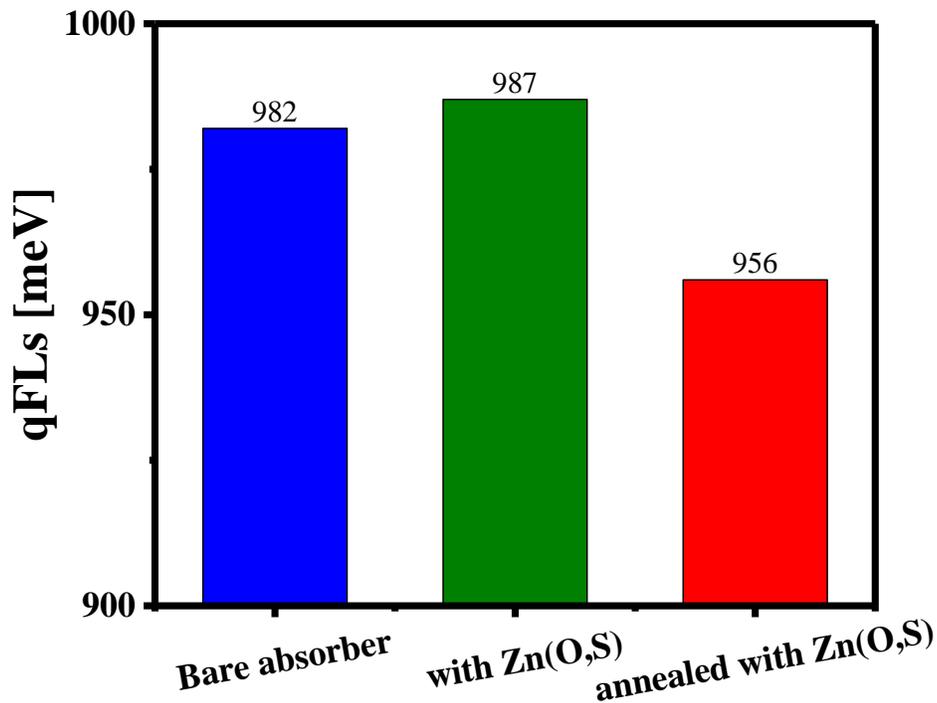

*Figure S1. Bar chart showing qFls values of CIGSu absorber without buffer (bare absorber), with chemical bath deposited Zn(O,S) buffer and with Zn(O,S) buffer annealed at 200 °C for 10 minutes.*

**Calibrated photoluminescence measurements**

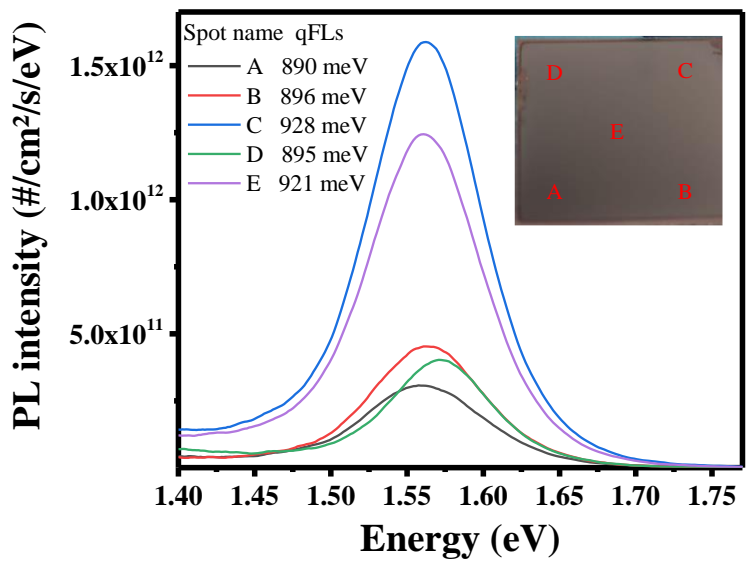

*Figure S2.* Exemplary calibrated PL spectra of one of the absorbers used to fabricate device. The qFLs is obtained by transforming these PL spectra using Planck's generalized law as explained in Ref[17, 28]. The PL signal varies in the sample at different spots and hence the qFLs.

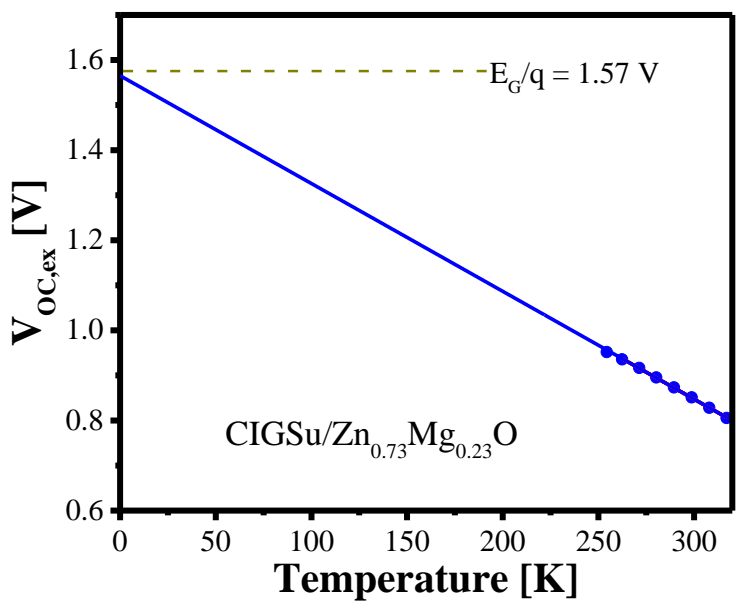

*Figure S3.* $V_{OC,ex}$ plotted as a function of temperature for $Cu(In,Ga)S_2/Zn_{0.73}Mg_{0.27}O/Al:ZnMgO/Al:ZnO$ device, the extrapolation of $V_{OC,ex}$ nearly is equal to $E_G/q$ suggesting bulk dominated device.

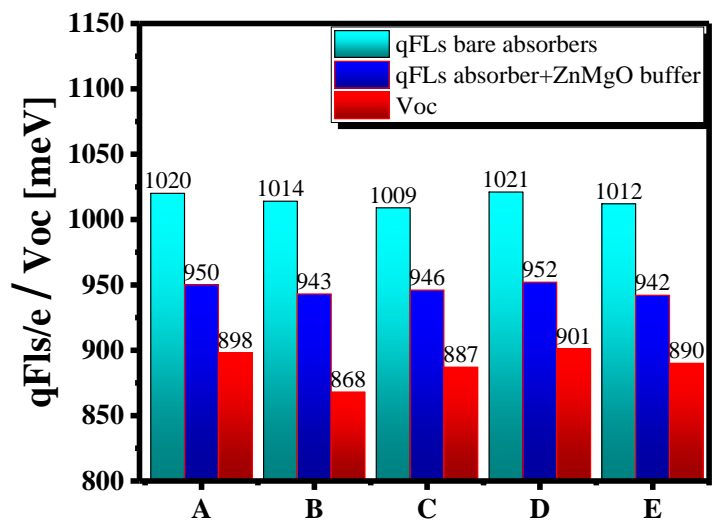

*Figure S4.* Exemplary bar chart of qFLs for the CIGSu absorber before and after ALD deposited $Zn_{0.73}Mg_{0.27}O$ buffer and the open-circuit voltage of the final device. A degradation in optoelectronic properties of the absorber has been observed whenever it is annealed during or after buffer deposition.

**J-V and EQE of best device:**

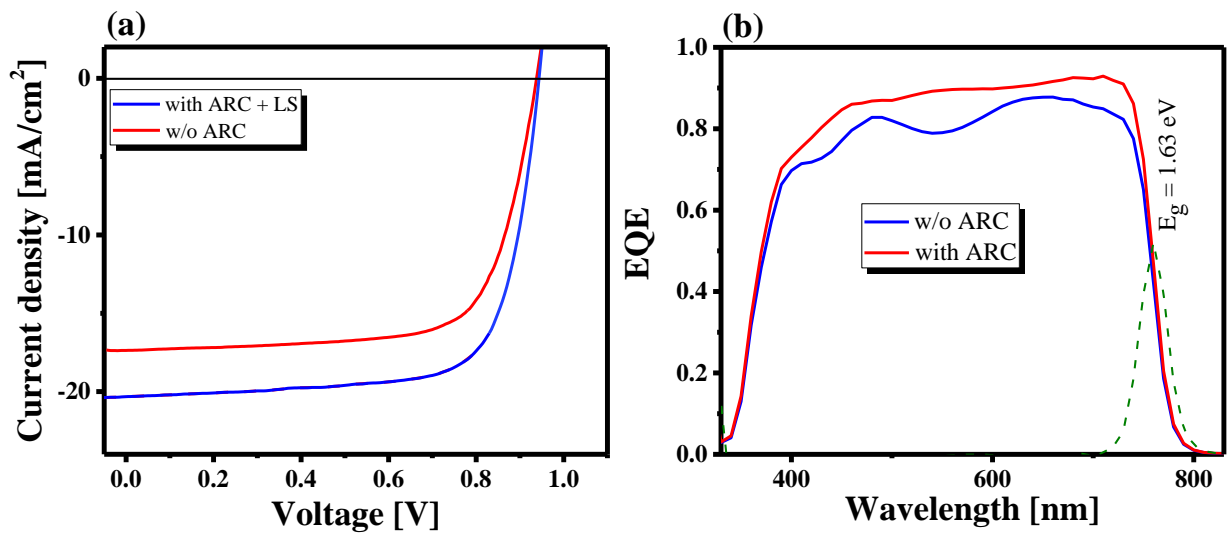

*Figure S5.* (a) J-V curve and (b) EQE curve of the CIGSu device having a PCE of 14.0% prepared with $Zn_{0.73}Mg_{0.27}O$ buffer layer and sputtered $Al:Zn_{0.73}Mg_{27}O$ second buffer layer with and without anti-reflective coating of $MgF_2$.

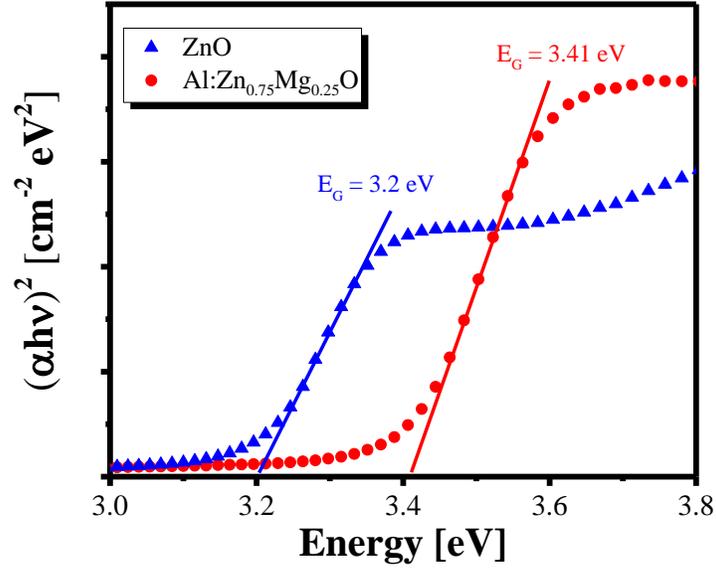

*Figure S6.* Tauc plot of sputtered ZnO and Al:ZnMgO. The extrapolation of linear part of the curve to x-axis gives the bandgap.

*Table S1.* SCAPS-1D numerical simulation parameters used to in this work. For achieving a value of $V_{OC,in}$ comparable to as observed in optical measurements, a deep defect level at 520 meV is introduced in the $Cu(In,Ga)S_2$ absorber layer.

| Parameter | $Cu(In,Ga)S_2$ | ZnMgO | Al:ZnMgO | ZnO | Al:ZnO |
|---|---|---|---|---|---|
| **Thickness (µm)** | 2.0 | 0.03 | 0.07 | 0.07 | 0.352 |
| **Band gap(eV)** | 1.57 | 3.5-3.7 | 3.5 | 3.3 | 3.45 |
| **Dielectric permittivity (relative)** | 10 | 10 | 10 | 10 | 10 |
| **Electron affinity(eV)** | 4.4 | 3.95-4.4 | 3.95-4.4 | 4.6 | 4.6 |
| **Electron mobility(cm²/Vs)** | 20 | 20 | 20 | 20 | 50 |
| **Hole mobility(cm²/Vs)** | 10 | 10 | 10 | 10 | 10 |
| **Doping(1/cm³)** | $1\times10^{15}$ | $1\times10^{17}$ | $1\times10^{17}$ | $1\times10^{17}$ | $5\times10^{19}$ |
| **Defect density(1/cm³)** | $2\times10^{16}$ acceptor 350meV from VBM | - | - | - | - |
| **Capture cross-section electrons (cm⁻²)** | $1\times10^{-14}$ | - | - | - | - |
| **Capture cross-section holes (cm⁻²)** | $1\times10^{-15}$ | - | - | - | - |

## ALD vs sputter-deposited i-layer

ALD deposited $Zn_{1-x}Mg_xO$ (x = 0.23) i-layer was also explored as a partner for the $Zn_{1-x}Mg_xO$ (x = 0.27) buffer layer. **Figure S7a** shows the J-V curves of CIGSu devices fabricated with ALD deposited $Zn_{1-x}Mg_xO$ (x = 0.27) buffer layer. One has an ALD deposited $Zn_{1-x}Mg_xO$ (x = 0.23) i-layer, the other one a sputtered Al:$Zn_{1-x}Mg_xO$ (x = 0.25) i-layer. In both cases sputtered Al:ZnO is still used as the TCO layer. As evident, the ALD deposited i-layer is inferior to sputtered i-layer. The device exhibits a higher interface $V_{OC}$ deficit (bar in the plot) and lower $J_{sc}$. Higher interface $V_{OC}$ deficit in the ALD deposited i-layer device might be caused by larger absorber degradation during i-layer deposition. The degradation in optoelectronic properties of CIGSu device, in general, is related to the heating of the device during or after the buffer deposition, which is independent of whether the buffer is a chemical bath deposited Zn(O,S) layer or an atomic layer deposited $Zn_{1-x}Mg_xO$ layer (see **Figure S1 and S4**). In the case of the ALD deposited i-layer process, the degradation is higher since it requires the absorber to be at 150 °C for extra three hours (to deposit ALD i-layer), as opposed to the process where the only buffer is deposited using ALD.

To understand the loss in $J_{sc}$ the EQE of the devices is measured and is plotted in **Figure S7b**. The EQE of the device drops as the wavelength decreases, causing a loss in $J_{sc}$ of the device. In comparison, this drop is much weaker in the case of sputtered Al:$Zn_{1-x}Mg_xO$ (x = 0.25) layer device. This loss indicates poor collection near the front interface. This could happen if the absorber is only weakly doped, *i.e.* the p = n point is deep in the absorber, and the interface has a high recombination probability for holes[26]. The photogenerated holes near the interface recombine while drifting towards the back contact. Consequently, the EQE and, therefore, the $J_{sc}$ drops. It is conceivable that the long ALD deposition time has reduced the absorber doping or increased the defects at the absorber-buffer interface. In summary, the sputtered Al:$Zn_{1-x}Mg_xO$ (x = 0.25) is the superior i-layer choice for CIGSu solar cells.

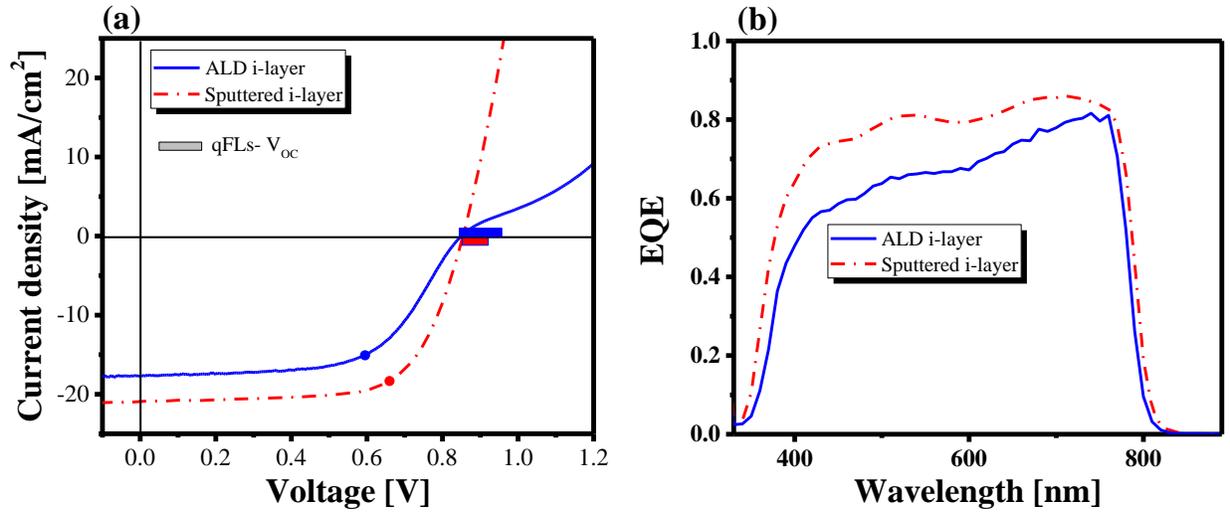

*Figure S7:* (a) Measured J-V curve and (b) EQE curve of the CIGSu device prepared with $Zn_{1-x}Mg_xO$ (x = 0.27) buffer layer with either ALD deposited $Zn_{1-x}Mg_xO$ (x = 0.23) i-layer or sputtered $Al:Zn_{1-x}Mg_xO$ (x = 0.25) i-layer. The bar in the J-V plot shows the deficit between qFLs/e (measured before buffer i-layer deposition) and $V_{OC}$.

# References

[1] P. Jackson *et al.*, "Properties of Cu (In, Ga) Se2 solar cells with new record efficiencies up to 21.7%," *Phys. Status. Solidi. R.,* vol. 9, no. 1, pp. 28-31, 2015, doi: https://doi.org/10.1002/pssr.201409520.

[2] P. Jackson *et al.*, "New world record efficiency for Cu (In, Ga) Se2 thin-film solar cells beyond 20%," *Prog. Photovolt.,* vol. 19, no. 7, pp. 894-897, 2011, doi: https://doi.org/10.1002/pip.1078.

[3] P. Jackson, R. Wuerz, D. Hariskos, E. Lotter, W. Witte, and M. Powalla, "Effects of heavy alkali elements in Cu(In,Ga)Se2 solar cells with efficiencies up to 22.6%," *Phys. Status Solidi RRL,* vol. 10, no. 8, pp. 583-586, 2016, doi: 10.1002/pssr.201600199.

[4] R. Carron *et al.*, "Advanced Alkali Treatments for High-Efficiency Cu (In, Ga) Se2 Solar Cells on Flexible Substrates," *Adv. Mater.,* vol. 9, no. 24, p. 1900408, 2019, doi: https://doi.org/10.1002/aenm.201900408.

[5] T. Kato, J.-L. Wu, Y. Hirai, H. Sugimoto, and V. Bermudez, "Record efficiency for thin-film polycrystalline solar cells up to 22.9% achieved by Cs-treated Cu (In,Ga)(Se,S)2," *IEEE J. Photovolt.,* vol. 9, no. 1, pp. 325-330, 2018, doi: https://doi.org/10.1109/JPHOTOV.2018.2882206.

[6] A. Miguel *et al.*, "Diode characteristics in state-of-the-art ZnO/CdS/Cu (In1-x Gax) Se2 solar cells," *Prog. Photovolt.: Res. Appl,* vol. 13, pp. 209-216, 2005.

[7] L. Weinhardt, C. Heske, E. Umbach, T. Niesen, S. Visbeck, and F. Karg, "Band alignment at the i-ZnO/CdS interface in Cu (In, Ga)(S, Se) 2 thin-film solar cells," *Appl. Phys. Lett.,* vol. 84, no. 16, pp. 3175-3177, 2004, doi: https://doi.org/10.1063/1.1704877.

[8] G. V. Rao, F. Säuberlich, and A. Klein, "Influence of Mg content on the band alignment at Cd S∕(Zn, Mg) O interfaces," *Appl. Phys. Lett.,* vol. 87, no. 3, p. 032101, 2005, doi: https://doi.org/10.1063/1.1995951.

[9] B. Tell, J. Shay, and H. J. P. r. B. Kasper, "Electrical Properties, Optical Properties, and Band Structure of CuGaS2 and CuInS2," *Phys. Rev. B.,* vol. 4, no. 8, p. 2463, 1971, doi: https://doi.org/10.1103/PhysRevB.4.2463.

[10] L. Weinhardt, O. Fuchs, D. Groß, G. Storch, E. Umbach, N. G. Dhere, A. A. Kadam, S. S. Kulkarni, and C. Heske, "Band alignment at the CdS∕Cu(In, Ga)S2 interface in thin-film solar cells," *Appl. Phys. Lett.,* vol. 86, no. 6, p. 062109, 2005, doi: https://doi.org/10.1063/1.1861958.

[11] S. Shukla *et al.*, "Over 15% efficient wide-band-gap Cu (In, Ga) S2 solar cell: Suppressing bulk and interface recombination through composition engineering," *Joule,* 2021, doi: https://doi.org/10.1016/j.joule.2021.05.004.

[12] C. Persson, C. Platzer-Björkman, J. Malmström, T. Törndahl, and M. Edoff, "Strong valence-band offset bowing of ZnO 1− x S x enhances p-type nitrogen doping of ZnO-like alloys," *Phys. Rev. Lett.,* vol. 97, no. 14, p. 146403, 2006, doi: https://doi.org/10.1103/PhysRevLett.97.146403.

[13] D. Kieven *et al.*, "Band alignment at sputtered ZnSx O1–x/Cu (In, Ga)(Se, S) 2 heterojunctions," *Phys. Status. Solidi. R.,* vol. 6, no. 7, pp. 294-296, 2012, doi: https://doi.org/10.1002/pssr.201206195.

[14] T. Minemoto, Y. Hashimoto, T. Satoh, T. Negami, H. Takakura, and Y. Hamakawa, "Cu (In, Ga) Se 2 solar cells with controlled conduction band offset of window/Cu (In, Ga) Se 2 layers," *J. Appl. Phys.,* vol. 89, no. 12, pp. 8327-8330, 2001, doi: https://doi.org/10.1063/1.1366655.

[15] H. Hiroi, Y. Iwata, S. Adachi, H. Sugimoto, and A. Yamada, "New World-Record Efficiency for Pure-Sulfide Cu(In,Ga)S$_2$; Thin-Film Solar Cell With Cd-Free Buffer Layer via KCN-Free Process," *IEEE J. Photovolt.,* vol. 6, no. 3, pp. 760-763, 2016, doi: https://doi.org/10.1109/JPHOTOV.2016.2537540.

[16] D. Regesch *et al.*, "Degradation and passivation of CuInSe 2," *Appl. Phys. Lett.,* vol. 101, no. 11, p. 112108, 2012, doi: https://doi.org/10.1063/1.4752165.

[17] F. Babbe, L. Choubrac, and S. Siebentritt, "Quasi Fermi level splitting of Cu-rich and Cu-poor Cu (In, Ga) Se2 absorber layers," *Appl. Phys. Lett.,* vol. 109, no. 8, p. 082105, 2016, doi: https://doi.org/10.1063/1.4961530.

[18] Y. Inoue, M. Hála, A. Steigert, R. Klenk, and S. Siebentritt, "Optimization of buffer layer/i-layer band alignment," in *2015 IEEE 42nd Photovoltaic Specialist Conference (PVSC)*, 2015: IEEE, pp. 1-5, doi: https://doi.org/10.1109/PVSC.2015.7355902.

[19] P. Wurfel, "The chemical potential of radiation," *Journal of Physics C: Solid State Physics,* vol. 15, no. 18, p. 3967, 1982, doi: https://doi.org/10.1088/0022-3719/15/18/012.

[20] M. Burgelman, P. Nollet, and S. Degrave, "Modelling polycrystalline semiconductor solar cells," *Thin Solid Films,* vol. 361-362, pp. 527-532, 2000/02/21/ 2000, doi: https://doi.org/10.1016/S0040-6090(99)00825-1.

[21] R. Scheer and H. Schock, "Thin film heterostructures," *Chalcogenide Photovoltaics,* pp. 9-127, 2011, doi: https://doi.org/10.1002/9783527633708.ch2.

[22] M. Sood *et al.*, "Near surface defects: Cause of deficit between internal and external open-circuit voltage in solar cells," *Prog. Photovolt.,* vol. 30, no. 3, pp. 263-275, 2021, doi: 10.1002/pip.3483.

[23] R. Scheer, "Activation energy of heterojunction diode currents in the limit of interface recombination," *J. Appl. Phys.,* vol. 105, no. 10, p. 104505, 2009, doi: https://doi.org/10.1063/1.3126523.

[24] W. Shockley and H. J. Queisser, "Detailed balance limit of efficiency of p-n junction solar cells," *J. Appl. Phys.,* vol. 32, no. 3, pp. 510-519, 1961, doi: https://doi.org/10.1063/1.1736034.

[25] A. Niemegeers and M. Burgelman, "Effects of the Au/CdTe back contact on IV and CV characteristics of Au/CdTe/CdS/TCO solar cells," *J. Appl. Phys.,* vol. 81, no. 6, pp. 2881-2886, 1997, doi: https://doi.org/10.1063/1.363946.

[26] R. Scheer and H. Schock, "Appendix A: Frequently Observed Anomalies," *Chalcogenide Photovoltaics,* pp. 305-314, 2011, doi: https://doi.org/10.1002/9783527633708.ch7.

[27] A. Niemegeers, S. Gillis, and M. Burgelman, "A user program for realistic simulation of polycrystalline heterojunction solar cells: SCAPS-1D," in *Proceedings of the 2nd World Conference on Photovoltaic Energy Conversion, JRC, European Commission, juli*, 1998, pp. 672-675, doi: https://biblio.ugent.be/publication/281948.

[28] T. Unold and L. Gütay, "Photoluminescence analysis of thin-film solar cells," *Advanced Characterization Techniques for Thin Film Solar Cells,* vol. 1, pp. 275-297, 2016, doi: https://doi.org/10.1002/9783527636280.ch7.